%% file: paper.tex
\documentclass[sigconf,authorversion]{acmart}

\usepackage{listings}
\usepackage[]{algorithm2e} % added <<<<<<<<<<<<<<

\usepackage{algorithmic} % added <<<<<<<<<<<<<<
\usepackage[nolist]{acronym}
\usepackage{adjustbox}
\usepackage{tabularx}
\usepackage{rotating}
\usepackage{multirow}
\usepackage{xcolor}
\usepackage{tcolorbox}
\usepackage{float}
\usepackage{lipsum}    % added
\usepackage{graphicx}  % added
\usepackage{subcaption}% added
\usepackage{todonotes}
\tcbuselibrary{skins,breakable}
\usepackage{booktabs}
\usetikzlibrary{arrows.meta}
\usepackage[edges]{forest}
\AtBeginDocument{\DeclareCaptionSubType{lstlisting}}
\usepackage{caption,subcaption}
\usepackage{array}

\usepackage{pdfrender}

\definecolor{codegreen}{rgb}{0,0.6,0}
\definecolor{codegray}{rgb}{0.5,0.5,0.5}
\definecolor{codepurple}{rgb}{0.58,0,0.82}
\definecolor{backcolour}{rgb}{0.95,0.95,0.92}

\lstdefinestyle{mystyle}{
    commentstyle=\color{codegreen},
    keywordstyle=\color{magenta},
    numberstyle=\tiny\color{codegray},
    stringstyle=\color{codepurple},
    basicstyle=\ttfamily\footnotesize,
    breakatwhitespace=false,         
    breaklines=true,                                     
    keepspaces=true,                 
    numbers=left,                    
    numbersep=5pt,                  
    showspaces=false,                
    showstringspaces=false,
    showtabs=false,                  
    tabsize=1
}

\newcommand{\draft}[1]{}
\renewcommand{\draft}[1]{{\color{blue} {#1}}}
\copyrightyear{2023} 
\acmYear{2023} 
\setcopyright{acmlicensed}\acmConference[SCORED '23]{Proceedings of the 2023 Workshop on Software Supply Chain Offensive Research and Ecosystem Defenses}{November 30, 2023}{Copenhagen, Denmark}
\acmBooktitle{Proceedings of the 2023 Workshop on Software Supply Chain Offensive Research and Ecosystem Defenses (SCORED '23), November 30, 2023, Copenhagen, Denmark}
\acmPrice{15.00}
\acmDOI{10.1145/3605770.3625212}
\acmISBN{979-8-4007-0263-1/23/11}
\begin{document}
\title{The Hitchhiker's Guide to Malicious Third-Party Dependencies} % TODO: replace with your title

\author{Piergiorgio Ladisa}
%\orcid{0000-0003-0850-4054}
\affiliation{%
  \institution{SAP Security Research}
  \streetaddress{805, avenue du Dr Maurice Donat}
  \city{Mougins}
  \country{France}
  \postcode{06250}
}
\affiliation{%
  \institution{University of Rennes 1/INRIA/IRISA}
  \streetaddress{263 avenue du Général Leclerc}
  \city{Rennes}
  \country{France}
  \postcode{35042}
}
\email{piergiorgio.ladisa@sap.com}
\email{piergiorgio.ladisa@irisa.fr}

\author{Merve Sahin}
\affiliation{%
  \institution{SAP Security Research}
  \streetaddress{805, avenue du Dr Maurice Donat}
  \city{Mougins}
  \country{France}
  \postcode{06250}
}
\email{merve.sahin@sap.com}

\author{Serena Elisa Ponta}
%\orcid{0000-0002-6208-4743}
\affiliation{%
  \institution{SAP Security Research}
  \streetaddress{805, avenue du Dr Maurice Donat}
  \city{Mougins}
  \country{France}
  \postcode{06250}
}
\email{serena.ponta@sap.com}

\author{Marco Rosa}
\affiliation{%
  \institution{SAP Security Research}
  \streetaddress{805, avenue du Dr Maurice Donat}
  \city{Mougins}
  \country{France}
  \postcode{06250}
}
\email{marco.rosa@sap.com}

\author{Matias Martinez}
%\orcid{0000-0002-2945-866X}
\affiliation{%
  \institution{Universitat Politècnica de Catalunya - Barcelona Tech}
  \streetaddress{Carrer de Jordi Girona, 31}
  \city{Barcelona}
  \country{Spain}
  \postcode{08034}
}
\email{matias.martinez@upc.edu}

\author{Olivier Barais}
%\orcid{0000-0002-4551-8562}
\affiliation{%
  \institution{University of Rennes 1/INRIA/IRISA}
  \streetaddress{263 avenue du Général Leclerc}
  \city{Rennes}
  \country{France}
  \postcode{35042}
}
\email{olivier.barais@irisa.fr}

\renewcommand{\shortauthors}{Piergiorgio Ladisa et al.}
%% No italics and no comma
%% If needed use a foot or author note to identify equal contribution

\begin{abstract}

%\draft{
    The increasing popularity of certain programming languages has spurred the creation of ecosystem-specific package repositories and package managers. 
%    These tools aim to streamline the use of open-source software for downstream users. 
    Such repositories (e.g., npm, PyPI) serve as public databases that users can query to retrieve packages for various functionalities, whereas package managers automatically handle dependency resolution and package installation on the client side.
    These mechanisms enhance software modularization and accelerate implementation. However, they have become a target for malicious actors seeking to propagate malware on a large scale. 
    
    In this work, we show how attackers can leverage capabilities of popular package managers and languages to achieve arbitrary code execution on victim machines, thereby realizing open-source software supply chain attacks. 
%    In particular, we investigate automated capabilities offered by popular package managers, which can potentially be exploited by attackers to trigger execution on victim machines. Then, we detail how attackers can achieve execution at installation, runtime, build time, or during tests.
%
%	In particular, we investigate capabilities offered by popular package managers that could be exploited by attackers to trigger execution on victim machines. 
    Based on the analysis of 7 ecosystems, we identify 3 install-time and 4 runtime techniques, and we provide recommendations describing how to reduce the risk when consuming third-party dependencies. We provide example implementations that demonstrate the identified techniques. Furthermore, we describe evasion strategies employed by attackers to circumvent detection mechanisms.
    %
%    We present proof-of-concept for each programming language, showcasing the triggering of execution through these automated features. Additionally, we discuss potential best practices and countermeasures that can be implemented to prevent such unintentional execution. By highlighting these practices, we aim to provide guidelines for developers to safeguard against the inadvertent triggering of execution and mitigate the risk of \ac{OSS} supply chain attacks.
% This paper investigates different techniques to inject malicious code into open
% source packages. 
%}

\end{abstract}

%% TODO: replace this section with code generated by the tool at https://dl.acm.org/ccs.cfm
%\begin{CCSXML}
%<ccs2012>
%<concept>
%<concept_id>10002978.10003029.10011703</concept_id>
%<concept_desc>Security and privacy~Usability in security and privacy</concept_desc>
%<concept_significance>500</concept_significance>
%</concept>
%</ccs2012>
%\end{CCSXML}

%\ccsdesc{Security and privacy~Use https://dl.acm.org/ccs.cfm to generate actual concepts section for your paper}
% -- end of section to replace with generated code

\begin{CCSXML}
    <ccs2012>
       <concept>
           <concept_id>10002978.10002997.10002998</concept_id>
           <concept_desc>Security and privacy~Malware and its mitigation</concept_desc>
           <concept_significance>500</concept_significance>
           </concept>
     </ccs2012>
\end{CCSXML}

\ccsdesc[500]{Security and privacy~Malware and its mitigation}

\keywords{Open-Source Security, Supply Chain Attacks, Malware Detection}

\maketitle

% ========== CONTENT  ========== %

\input{sections/intro}

\input{sections/supply-chain}

\input{sections/approach}

\input{sections/rq1}

\input{sections/rq2}

\input{sections/relwork}

\input{sections/conclusions}

% \section*{Acknowledgments}
%	This work was partly supported by the EC within the H2020 under grant agreement XXX (NAME).
\hfill \break
%%%%%%%%%%%%%%%%%%%%%%%%%%%%%%%%%%%%%%%%%%%%%%%%%%%%%%%%%%%%%%%%%%%%%%%%%%%%%%%%
\small\noindent\textbf{Acknowledgements.}
%\section*{Acknowledgments}
%{\skip0=\baselineskip\small\baselineskip=\skip0 
We thank the reviewers for their constructive feedback, which has greatly contributed to the improvement of our work.
This work is partly funded by EU grants No. 952647 (AssureMOSS) and No. 101120393 (Sec4AI4Sec).
\normalsize

% ========== BIBLIOGRAPHY ========== %

\bibliographystyle{ACM-Reference-Format}
\balance
\bibliography{references}

\begin{acronym}[TDMA]
    \acro{AV}{Anti Virus}
    \acro{AVs}{Anti Viruses}
    \acro{ACE}{Arbitrary Code Execution}
    \acro{CTI}{Cyber Threat Intelligence}
    \acro{C2}{Command and Control}
    \acro{CPG}{Code Property Graph}
    \acro{TUF}{The Update Framework}
    \acro{PKI}{Public Key Infrastructure}
    \acro{CI}{Continuous Integration}
    \acro{CD}{Continuous Delivery}
    \acro{UI}{User Interface}
    \acro{VCS}{Versioning Control System}
    \acro{VMs}{Virtual Machines}
    \acro{VA}{Vulnerability Assessment}
    \acro{VCS}{Version Control System}
    \acro{SCM}{Source Control Management}
    \acro{IAM}{Identity Access Management}
    \acro{CDN}{Content Delivery Network}
    \acro{UX}{User eXperience}
    \acro{SLR}{Systematic Literature Review}
    \acro{SE}{Social Engineering}
    \acro{MITM}{Man-In-The-Middle}
    \acro{SBOM}{Software Bill of Materials}
    \acro{MFA}{Multi-Factor Authentication}
    \acro{AST}{Abstract Syntax Tree}
    \acro{RASP}{Runtime Application Self-Protection}
    \acro{OS}{Operating System}
    \acro{OSS}{Open-Source Software}
    \acro{TARA}{Threat Assessment and Remediation Analysis}
    \acro{CAPEC}{Common Attack Pattern Enumeration and Classification}
    \acro{DoS}{Denial of Service}
    \acro{SCA}{Software Composition Analysis}
    \acro{SLSA}{Supply-chain Levels for Software Artifacts}
    \acro{SDLC}{Software Development Life-Cycle}
    \acro{ICT}{Information and Communication Technologies}
    \acro{C-SCRM}{Cyber Supply Chain Risk Management}
    \acro{DDC}{Diverse Double-Compiling}
    \acro{OSINT}{Open Source Intelligence}
    \acro{U/C}{Utility-to-Cost}
    \acro{LOC}{Lines Of Code}
    \acro{JVM}{Java Virtual Machine}
    \acro{LFI}{Local File Inclusion}
    \acro{PII}{Personally Identifying Information}
    \acro{RCE}{Remote Code Execution}
    \acro{JIT}{Just-In-Time}
    \acro{DoS}{Denial of Service}
    \acro{JAR}{Java Archive}
    \acro{AOT}{Ahead-Of-Time}
    \acro{DT}{Decision Tree }
    \acro{RF}{Random Forest }
    \acro{XGBoost}{eXtreme Gradient Boosting}
    \acro{BO}{Bayesian Optimizer}
    \acro{GL}{Generalization Language}
    \acro{GL3}{}
    \acro{GL4}{}
    \acro{ML}{Machine Learning}
    \acro{BKC}{Backstabber's Knife Collection}
    \acro{KS}{Kolmogrov-Smirnov}
    \acro{PoC}{Proof-of-Concept}
    \acro{TTPs}{Tactics, Techniques, and Procedure(s)}
    %\acro{AST}{Abstract Syntax Tree}
\end{acronym}

% ========== APPENDIX ========== %
% \appendix
% \input{sections/appendix}

\end{document}

%% file: sections/intro.tex
\section{Introduction}

Software modularization is a fundamental practice in modern software development, as it enables the division of complex systems into more manageable components and promotes software reusability. In this context, \ac{OSS} plays a central role by offering a wide range of pre-built and reusable modules that developers can leverage to enhance productivity. 
To streamline the use of \ac{OSS}, package repositories and managers for specific programming languages have been established. These repositories serve as centralized databases where developers can easily discover, download, install, and manage open-source modules.
On the client side, package managers are tools that assist downstream users in automatically handling the resolution and installation of required packages, including their dependencies. 
These mechanisms are popular among developers, but the full automation they provide can involve potentially risky processes that remain transparent to the users.
In fact, package repositories have become a fruitful target for attackers seeking to propagate malware on a large scale~\cite{pytorchSC2023,lolipop2023,iconburst2022}.
\ac{OSS} supply chain attacks are an increasing trend and demonstrated to happen through multiple attack vectors~\cite{ladisa2022taxonomy,ohm2020backstabbers}. The severity of this issue is highlighted by the suspension in May 2023 of new user registrations and package uploads on PyPI due to a significant increase in malicious activities~\cite{pythonPyPIUser}.

Given the significance of \ac{OSS}, which can account for up to 98\% of codebases~\cite{nagle2022census}, and the widespread adoption of package management practices by both private and public organizations, enhancing the security of the software supply chain has become imperative. This is necessary to protect the interests of the community and ensure national security~\cite{sonatypeAnnualState, whitehouseExecutiveOrder, enisaThreatLandscape2022}.

% Considering the importance of open-source software (which can constitute up to the 98\% of codebases~\cite{nagle2022census}) and package management practices for both private and public organizations, improving the security of the software supply chain has become essential to safeguard the interests of the community and ensure national security~\cite{sonatypeAnnualState, whitehouseExecutiveOrder, enisaThreatLandscape2022}. 

%In this work, we adopt the offensive perspective of an attacker and investigate the methods by which achieving \ac{ACE} is possible through the distribution of malicious packages to downstream users. 
In this work, we investigate how to achieve \ac{ACE} through the distribution of malicious packages to downstream users. 
Moreover, we delve into the evasion techniques that attackers may employ to circumvent detection measures.
We set out to answer the following research questions.

%\todo{research questions are a bit verbose (repeating the same thing for install/run/build/test. we can simplify or make a twist to make questions more interesting. }

%\textbf{RQ1:} What features provided by popular package managers allow to third-party dependencies to achieve \ac{ACE} on downstream projects during the package life-cycle phases (i.e., install, runtime)?\todo
\textbf{RQ1:} How can 3rd-party dependencies, distributed via package managers, achieve \ac{ACE} on downstream projects during the third-party package life-cycle phases (i.e., install, runtime)?

\textbf{RQ2:} What are the strategies adopted by attackers to evade the detection of malicious code?

% \textbf{RQ3:} What are the possible recommendations for downstream users and security analysts to mitigate the risks associated with consuming 3rd-party dependencies?

% \textbf{RQ2:} What are the ways to achieve code execution at build-time?

% \textbf{RQ3:} What are the best locations to hide malicious code to increase the chances of execution?

% \textbf{RQ4:} What are the ways to achieve code execution during test-time?

To answer these questions, we first explore the functionalities of package managers for popular languages that allow to trigger execution at install-time and runtime.
Then, we identify where attackers can hide malicious code to increase the chances of achieving its execution when the downstream project is built, tested, or run. %at runtime or during test.
Based on the analysis of 7 ecosystems, we identify 3 install-time techniques and 4 runtime techniques.
Finally, we describe the various techniques employed by attackers to evade detection.

For the identified techniques, we provide practical guidance to downstream users on how to prevent \ac{ACE} and to security analysts on how to analyze malicious packages.
In addition, we release the source code of example implementations\footnote{\url{https://github.com/SAP-samples/risk-explorer-execution-pocs}} available for the covered programming languages and their reference package manager, i.e., Python (pip), JavaScript (npm), Ruby (gem), PHP (composer), Rust (cargo), Go (go), and Java (mvn). This release aims to support fellow researchers in devising protective measures against such threats. We believe this may also help penetration testers to test whether the offensive techniques explored in this work may be achieved within their organizations.

%\smallskip\noindent
%\draft{\textbf{Outline} 
The remainder of the paper is organized as follows.
Section~\ref{sec:oss-ssc-attacks} provides background information on 3rd-party dependencies and \ac{OSS} supply chain attacks.
Section~\ref{sec:approach} presents our approach.
%Section~\ref{sec:rq1} answers to RQ1, showing several supply chain attacks strategies, organized per time of execution triggering the malicious code, and providing practical examples for several programming languages for each of these strategies.
Section~\ref{sec:rq1} answers to RQ1, showing techniques to achieve \ac{ACE} at install-time and runtime. %, and provides practical examples for several programming languages.
%install-time, build-time, runtime and test time
Section~\ref{sec:rq2} answers to RQ2, describing evasion techniques. % that can be used to better hide malicious code. %, and thus make malicious packages aiming at supply chain attacks more difficult to discover.
%Section~\ref{sec:ethics} presents some ethical considerations.
Section~\ref{sec:related-works} discusses related work.
Finally, Section~\ref{sec:conclusion} presents our conclusions.
%}

%% file: sections/supply-chain.tex
\section{Background}\label{sec:oss-ssc-attacks}

\begin{figure}[tp!]
	\centering
	\includegraphics[width=.48\textwidth]{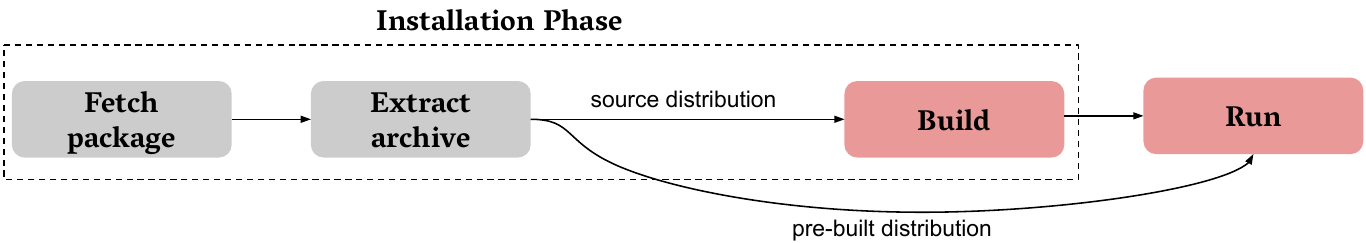}
	\caption{Lifecycle of a 3rd-party dependency (both direct and transitive) from the perspective of a downstream project.}
	\label{fig:pkg-lifecycle}
\end{figure}

Modern software development extensively relies on 3rd-party dependencies. Dependencies are classified as \textit{direct} when they are explicitly declared within a downstream application. In turn, direct dependencies may themselves have additional dependencies, referred to as \textit{transitive} dependencies.
Figure~\ref{fig:pkg-lifecycle} shows the lifecycle of a 3rd-party dependency (both direct and transitive) from the perspective of a downstream project.
%From the perspective of the downstream project, the lifecycle of a 3rd-party dependency (both direct and transitive) can be described as in Figure~\ref{fig:pkg-lifecycle}. 
%
We can distinguish two main phases: \textit{install} and \textit{run}.
During the installation phase, the package manager on the client side fetches the dependency (i.e., a package) from the package repository and extracts it locally. If the package distribution is of \textit{source} type (i.e., it contains only the source code), it is built for the target architecture of the downstream users. On the other hand, if the distribution is pre-built, the retrieved package can be used directly.
Once the 3rd-party dependency is installed, it can be run within downstream projects as part of the main application or during the tests (in case of \textit{test dependencies}).

\ac{OSS} supply chain attacks encompass the insertion of malicious code into open-source components, enabling the distribution of malware to downstream consumers~\cite{ohm2020backstabbers,ladisa2022taxonomy}. 
%These attacks can be conducted through diverse attack vectors, as documented in~\cite{ladisa2022taxonomy}. Broadly speaking, attackers may opt to: \textit{develop and promote a malicious package from scratch}, \textit{create name confusion with a legitimate package}, or \textit{compromise a legitimate package} (i.e., injecting malicious code into the source code, during the build process, or within the package repository).
%
To execute a successful \ac{OSS} supply chain attack, the attacker must fulfill mainly three requirements~\cite{sejfia2022practical}: (1) make a malicious package accessible to downstream users; (2) ensure that the downstream users actively engage with the malicious package (e.g., by installing it); (3) ensure that the downstream users eventually execute the malicious code contained within the package. 
While the taxonomy of attacks in~\cite{ladisa2022taxonomy} mostly covers the first two steps of this process, our work focuses on the last step, i.e., ensuring that downstream users eventually execute the malicious code.
%The malicious code can have different purposes, e.g., exfiltrating sensitive information, spawning a reverse shell, running cyrptominers~\cite{ohm2020backstabbers}.  %\todo{seems out of scope here}

% Once an attacker is able to inject code into a legitimate package, or publish their own malicious package in a repository, the downstream users are likely to download and install this package via package managers, and later import/use the relevant libraries to code their applications. Our study focuses on this process, understanding how package managers and programming language relevant features enable the execution of malicious code at different parts of development lifecycle??

Throughout this paper, we utilize the terminology \ac{TTPs} as defined by the MITRE ATT\&CK framework \cite{mitre_attack}.
In our context, tactics refer to the main objectives of an attack, which in our work entails achieving \ac{ACE} during installation time and runtime. Techniques encompass the specific methods employed by attackers, while procedures outline the practical implementation details followed by attackers.

%% file: sections/approach.tex
\section{Approach}\label{sec:approach}

\begin{figure}[t!p]
	\centering
	\includegraphics[width=.48\textwidth]{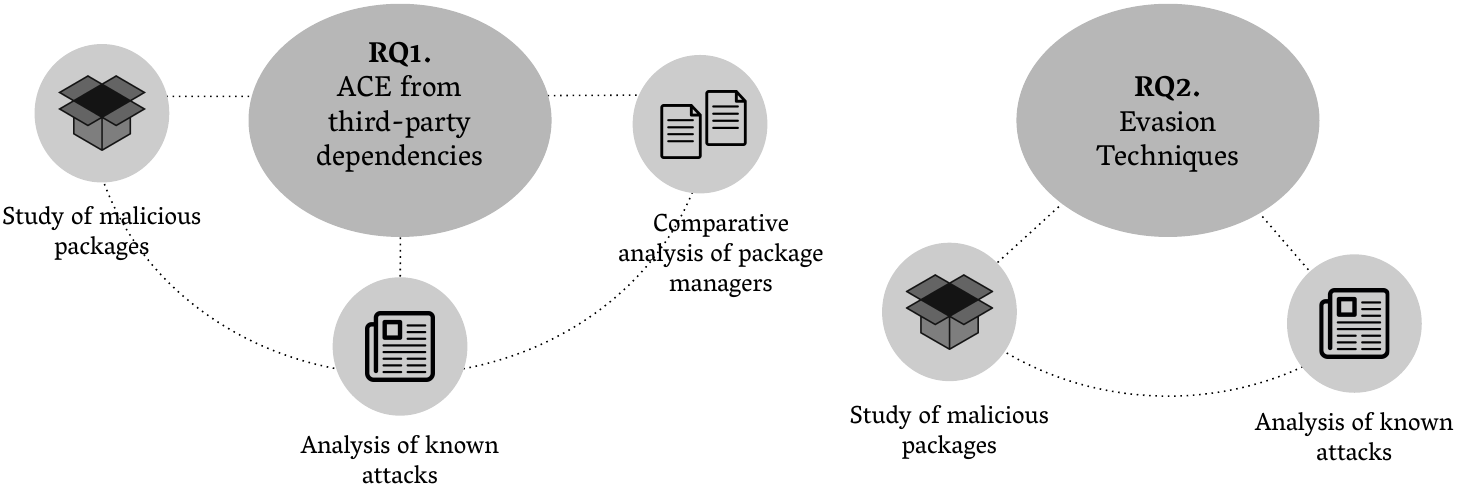}
	\caption{Approach followed to answer the research questions.}
	\label{fig:approach}
\end{figure}

To address both RQ1 and RQ2, we adopt the approach in Figure~\ref{fig:approach} and we rely on two primary data sources.

The first is the \ac{BKC}~\cite{dasfreakBackstabbersKnife}, a dataset containing malicious packages from past attacks. This dataset spans various programming languages, i.e., JavaScript (npm), Python (pip), PHP (composer), Ruby (gem), and Java (mvn). We conduct a comprehensive manual analysis of these packages. 

The second source is the Risk Explorer~\cite{10.1145/3560835.3564546}, which offers a comprehensive taxonomy and an extensive dataset of references related to \ac{OSS} supply chain attacks. In this instance, we analyze grey literature materials to extract technical details from previous attacks, identifying any novel or advanced techniques that may not be present in the \ac{BKC}.

% Our approach begins with a manual examination of packages within the \ac{BKC}~\cite{dasfreakBackstabbersKnife}. This investigation aims to uncover the mechanisms that trigger the execution of malicious code and the evasion techniques employed. Notably, the \ac{BKC} encompasses packages from diverse programming languages, i.e., JavaScript (npm), Python (pip), PHP (composer), Ruby (gem), and Java (mvn).
% Subsequently, we delve into the grey literature available via the Risk Explorer~\cite{10.1145/3560835.3564546}. This analysis focuses on extracting technical insights from past attacks, identifying any innovative or advanced techniques that may not be present in the \ac{BKC}.

To address RQ1, we first study the known attacks and malicious packages in~\cite{10.1145/3560835.3564546} and~\cite{dasfreakBackstabbersKnife} to identify the underlying root causes to achieve \ac{ACE}.
This involved examining the features provided by the language or the package managers that were exploited to achieve \ac{ACE} by means of 3rd-party dependencies.
To evaluate the presence and similarity of functionalities across diverse programming languages, we conduct a \textit{comparative analysis of package managers' functionalities}. This examination involves scrutinizing the documentation of package managers associated with the selected programming languages to determine if similar functionalities exist and if they pose similar risks.
In expanding this comparative analysis beyond the languages covered in~\cite{10.1145/3560835.3564546} and~\cite{dasfreakBackstabbersKnife}, we include Go (go), chosen for its explicit focus on addressing supply chain attacks~\cite{goSupplyChain}, and Rust (cargo), due to its increasing popularity~\cite{stackoverflowStackOverflow, zdnetLinusTorvalds}.
Upon identifying functionalities that could potentially lead to \ac{ACE}, we proceed to develop a set of runnable implementations. These implementations provide minimal examples of each specific exploitation technique, showcasing the root cause of the issues and the different code locations susceptible to housing malicious content.

To address RQ2, we build upon the evasion techniques identified in~\cite{dasfreakBackstabbersKnife} and supplement them with evasion strategies documented in the grey literature references from~\cite{10.1145/3560835.3564546}.
In this pursuit, we also refer to the literature on code obfuscation~\cite{10.1145/2886012,1027797, xu2017secure,collberg1997taxonomy} to identify additional evasion techniques that share similarities with those already exploited but have not been observed in the wild (to the best of our knowledge). 
We adopt the categorization of evasion techniques proposed by Schrittwieser et al.~\cite{10.1145/2886012}, which includes \textit{data obfuscation}, \textit{static code transformation}, and \textit{dynamic code transformation}.

% Finally, we differentiate between techniques to
% obfuscate source code, and strings in particular. 
% %TODO: the last sentence should be updated if we can adopt a categorization
% %method from previous work.

Our work presents an extensive overview of both observed and potential \ac{ACE} mechanisms and evasion tactics. This information can help defenders in recognizing malicious packages and devising mitigation strategies within the software supply chain. 
However, we do not claim to be exhaustive, as there could be other (unknown) techniques for triggering the execution of malicious code and evading detection.

% Our work provides an extensive overview of the ACE mechanisms and evasion
% techniques that are observed in the wild, and also explores other mechanisms and
% techniques that are technically feasible, thus potentially exploitable.  Such an
% overview can help defenders to better identify the malicious packages, and to
% support mitigation activities in the software supply chain.  However, we do not
% claim to be exhaustive, as there may be other (unknown) viable techniques to
% create malicious packages and evade detection.

%% file: sections/rq1.tex
\section{RQ1: Arbitrary Code Execution Strategies} \label{sec:rq1}

In this section we answer RQ1, describing 
techniques that 3rd-party dependencies may employ to attain \ac{ACE} when they are installed or run in the context of downstream projects.
Building on Ohm et al.'s analysis~\cite{ohm2020backstabbers}, these techniques are enhanced through insights from grey literature and package manager documentation.

% Figure~\ref{fig:pkg-lifecycle} represents the lifecycle of a dependency (both direct and transitive) from the perspective of a downstream project. 

% During the installation phase, the package manager on the client side fetches the the resource from the package repository and extracts it locally. If the package distribution is of the \textit{source} type, meaning it contains only the source code, it needs to be built for the target architecture of the downstream users before it can be used. On the other hand, if the distribution is pre-built, it is already prepared for immediate usage.

% Once the 3rd-party dependency is installed, it can be run on downstream projects as part of the main application or during the tests (i.e., the case of \textit{test dependencies}).

% When using 3rd-party dependencies in tests, they can employ the techniques described for the runtime cases to achieve \ac{ACE} (e.g., import-time case).

% As an additional remark, attackers may ship packages with malicious code in the test routines. This is out of our scope since our focus is on 3rd-party dependencies that would execute code when testing our project (and not their code). However, if victims run tests of the dependencies, they should ensure that they do not contain malicious code.

Table~\ref{tab:ace-techniques} provides a summary of the 7 techniques we have identified for achieving \ac{ACE}. Table~\ref{tab:lang-comparison} maps these techniques to the selected languages/package managers and indicates whether each technique is applicable in a particular ecosystem. We provide example implementations for all these techniques.

\begin{table*}[!tp]
	% Green: \cellcolor[HTML]{98FB98}
	% Yellow: \cellcolor[HTML]{FBB917}
	% Red: \cellcolor[HTML]{FF0000}

	\centering
	\begin{adjustbox}{angle=0}
		\begin{tabular}{r|ccc|}
			%\toprule
			
			%\rowcolor[HTML]{9B9B9B} 
			\multicolumn{3}{c}{\textbf{How to achieve \acl{ACE}}} \\ \cmidrule(lr){1-3}  
			\multicolumn{1}{c}{\textbf{Package Life-Cycle Phase}} & \multicolumn{1}{c}{\textbf{Technique}} & \multicolumn{1}{c}{\textbf{Example}} \\ \cmidrule(lr){1-1} \cmidrule(lr){2-2}  \cmidrule(lr){3-3}
			
			\multicolumn{1}{p{2cm}}{(\textbf{I}) Install-time} & \multicolumn{1}{p{8cm}}{(\textbf{I1}) Run command/scripts leveraging install-hooks~\cite{ohm2020backstabbers}} & \multicolumn{1}{p{5cm}}{In JavaScript (npm), insert a shell command as value of the key \texttt{pre-install} in \textit{package.json} (cf. Listing~\ref{lst:js-poc1}).}  \\
			
			\multicolumn{1}{p{2cm}}{}& \multicolumn{1}{p{8cm}}{(\textbf{I2}) Run code in build script~\cite{ohm2020backstabbers}} & \multicolumn{1}{p{5cm}}{In Python, insert code in \textit{setup.py} (cf. Listing~\ref{lst:py-poc1}).}  \\
			
			\multicolumn{1}{p{2cm}}{}& \multicolumn{1}{p{8cm}}{(\textbf{I3}) Run code in build extension(s) } & \multicolumn{1}{p{5cm}}{In Ruby (gem), insert code in build extension, such as the \textit{extconf.rb} file (cf. Listing~\ref{lst:ruby-poc}).}  \\
			
			% \multicolumn{1}{p{2cm}}{(\textbf{B}) Build-time} & \multicolumn{1}{p{8cm}}{(\textbf{B1}) Insert code in build extension} & \multicolumn{1}{p{5cm}}{As described in I1.3 (cf. Listing~\ref{lst:ruby-poc}).}  \\
			
			% \multicolumn{1}{p{2cm}}{} & \multicolumn{1}{p{8cm}}{(\textbf{B2}) Insert code in build script} & \multicolumn{1}{p{5cm}}{Insert code in \textit{build.rs} for Rust (cf. Listing~\ref{lst:rust-poc}).}  \\
			
			\multicolumn{1}{p{2cm}}{(\textbf{R}) Runtime} & \multicolumn{1}{p{8cm}}{(\textbf{R1}) Insert code in methods/scripts executed when importing a module~\cite{ohm2020backstabbers}} & \multicolumn{1}{p{5cm}}{In Python, insert code in \textit{\_\_init\_\_.py} file (cf. Section~\ref{sec:runtime-exec}).}  \\

			\multicolumn{1}{p{2cm}}{} & \multicolumn{1}{p{8cm}}{(\textbf{R2}) Insert code in commonly-used methods} & \multicolumn{1}{p{5cm}}{In Python, insert code in \texttt{setup()} method of \texttt{distutil.core}.}  \\
			
			\multicolumn{1}{p{2cm}}{} & \multicolumn{1}{p{8cm}}{(\textbf{R3}) Insert code in constructor methods (of popular classes)} & \multicolumn{1}{p{5cm}}{In Python, insert code in \texttt{Dataframe()} constructor method of package typosquatting \texttt{pandas} package.}  \\

			% \multicolumn{1}{p{2cm}}{} & \multicolumn{1}{p{8cm}}{(\textbf{R4}) Override popular methods, e.g., using monkey-patch~\cite{ohm2020backstabbers} or function hooking~\cite{bouMonkeyPatching}.} & \multicolumn{1}{p{5cm}}{Overwrite the \texttt{setup()} method of \texttt{distutil.core} package in Python.}  \\

			\multicolumn{1}{p{2cm}}{} & \multicolumn{1}{p{8cm}}{(\textbf{R4}) Run code of 3rd-party dependency as build plugin}  & \multicolumn{1}{p{5cm}}{In Java (mvn), insert code in maven plugin that is executed when building downstream project.}\\
			
			%\multicolumn{1}{p{2cm}}{} & \multicolumn{1}{p{8cm}}{(\textbf{R1.5}) Replace popular methods using function hooking~\cite{bouMonkeyPatching}} & \multicolumn{1}{p{5cm}}{}  \\
			
			%\multicolumn{1}{p{2cm}}{(\textbf{T}) Test-time} & \multicolumn{1}{p{8cm}}{(\textbf{T1}) Poison test dependency} & \multicolumn{1}{p{5cm}}{TODO}  \\
			
			%\bottomrule
		\end{tabular}
	\end{adjustbox}
	\caption{ Techniques that a 3rd-party dependency may use to achieve \ac{ACE} on downstream during its lifecycle.}
	\label{tab:ace-techniques}
\end{table*}

\begin{table}[!tp]
	% Green: \cellcolor[HTML]{98FB98}
	% Yellow: \cellcolor[HTML]{FBB917}
	% Red: \cellcolor[HTML]{FF0000}
	
	\centering
	\begin{adjustbox}{angle=0}
		\begin{tabular}{r|llllllllllll|}
			\toprule
			
			%\rowcolor[HTML]{9B9B9B} 
			\multicolumn{1}{l}{} & \multicolumn{8}{c}{\textbf{ACE Technique(s)}} \\ 
			\multicolumn{1}{l}{} & \multicolumn{3}{c}{\textbf{Install-time}} & \multicolumn{5}{c}{\textbf{Runtime}} \\
			\multicolumn{1}{c}{\textbf{Ecosystem}} & \multicolumn{1}{c}{\textbf{I1}} & \multicolumn{1}{c}{\textbf{I2}} & \multicolumn{1}{c}{\textbf{I3}} & \multicolumn{1}{c}{\textbf{R1}} & \multicolumn{1}{c}{\textbf{R2}} & \multicolumn{1}{c}{\textbf{R3}} & \multicolumn{1}{c}{\textbf{R4}}  \\ \cmidrule(lr){1-1} \cmidrule(lr){2-4} \cmidrule(lr){5-9} 
			
			\multicolumn{1}{c}{JavaScript (npm)} & \multicolumn{1}{c}{\checkmark} & \multicolumn{1}{c}{} &
			\multicolumn{1}{c}{} & \multicolumn{1}{c}{\checkmark} &
			\multicolumn{1}{c}{\checkmark} & \multicolumn{1}{c}{\checkmark} &
			%\multicolumn{1}{c}{\checkmark}&
			
			\\
			
			\multicolumn{1}{c}{Python (pip)}  & \multicolumn{1}{c}{} & \multicolumn{1}{c}{\checkmark} &
			\multicolumn{1}{c}{} & \multicolumn{1}{c}{\checkmark} &
			\multicolumn{1}{c}{\checkmark} & \multicolumn{1}{c}{\checkmark} &
			%\multicolumn{1}{c}{\checkmark}&
			\\
			
			\multicolumn{1}{c}{PHP (composer)} & \multicolumn{1}{c}{\checkmark} & \multicolumn{1}{c}{} &
			\multicolumn{1}{c}{} & \multicolumn{1}{c}{} &
			\multicolumn{1}{c}{\checkmark} & \multicolumn{1}{c}{\checkmark} &
			\multicolumn{1}{c}{}&
			\\
			
			\multicolumn{1}{c}{Ruby (gem)}  & \multicolumn{1}{c}{} & \multicolumn{1}{c}{} &
			\multicolumn{1}{c}{\checkmark}  & \multicolumn{1}{c}{\checkmark} &
			\multicolumn{1}{c}{\checkmark} & \multicolumn{1}{c}{\checkmark} &
			%\multicolumn{1}{c}{\checkmark}&
			\\
			
			\multicolumn{1}{c}{Rust (cargo)} & \multicolumn{1}{c}{} & \multicolumn{1}{c}{\checkmark} &
			\multicolumn{1}{c}{}  & \multicolumn{1}{c}{} &
			\multicolumn{1}{c}{\checkmark} & \multicolumn{1}{c}{\checkmark} &
			%\multicolumn{1}{c}{\checkmark}  
			\\
			
			\multicolumn{1}{c}{Go (go)}  & \multicolumn{1}{c}{} & \multicolumn{1}{c}{} &
			\multicolumn{1}{c}{} & \multicolumn{1}{c}{\checkmark} &
			\multicolumn{1}{c}{\checkmark} & \multicolumn{1}{c}{\checkmark} &
			%\multicolumn{1}{c}{\checkmark}  
			\\
			
			\multicolumn{1}{c}{Java (mvn)}  & \multicolumn{1}{c}{} & \multicolumn{1}{c}{} &
			\multicolumn{1}{c}{} & \multicolumn{1}{c}{} &
			\multicolumn{1}{c}{\checkmark} & \multicolumn{1}{c}{\checkmark} &
			\multicolumn{1}{c}{\checkmark} % &
			%\multicolumn{1}{c}{\checkmark}
			\\

			\bottomrule
		\end{tabular}
	\end{adjustbox}
	\caption{Comparative analysis about applicability of available techniques per each language (and related package manager). Empty cells in the table indicate that a particular technique is not applicable to a specific language. We provide implementations for all techniques for each programming language.}
	\label{tab:lang-comparison}
\end{table}

\subsection{(I) Install-Time Execution}\label{sec:install-time}

We identify 3 techniques to achieve \ac{ACE} when downstream users/projects install a 3rd-party dependency using popular package managers (i.e., npm, composer, pip, and gem).
%In this section we describe how to achieve \ac{ACE} in the context of popular package managers (i.e., npm, composer, pip, and gem) when installing a 3rd-party dependency.

\subsubsection*{(I1) Run command/scripts leveraging install-hooks~\cite{ohm2020backstabbers}}
This technique involves the execution of code by hooking the install process of 3rd-party dependencies in different stages, using specific keywords that package managers may provide.

\textit{JavaScript (npm).}
%\noindent\textit{JavaScript (npm).}
%In Node.js, the \textit{package.json} file contains both metadata information (e.g., name, version) and the list of dependencies related to a project~\cite{npmjsPackagejsonDocs}. Thus, this file is processed by the package manager to install and resolve the dependencies for a specific package. 
%The \textit{package.json} file also provides installation hooks through the \textit{scripts} property, which is a dictionary where the keys represent lifecycle phases (e.g., \textit{pre-install}), indicating when the scripts in the corresponding values should be executed~\cite{npmjsPackagejsonDocs}. 
%
%
In Node.js, the \textit{package.json} file contains both metadata information (e.g., name, version) and the list of dependencies related to a project~\cite{npmjsPackagejsonDocs}. It also provides installation hooks through the \textit{scripts} property: keys for different lifecycle phases (e.g., \textit{pre-install}) can be used to trigger the execution of the scripts provided as values~\cite{npmjsPackagejsonDocs}. 
The \textit{package.json} file is processed by the package manager to install and resolve the dependencies for a specific package. 
During the installation process, scripts are executed according to their definition in the corresponding property to perform additional actions (e.g., to compile code)~\cite{10.1145/3488932.3523262}.

\noindent\textbf{Procedure.} To achieve \ac{ACE} when installing a package through \texttt{npm install}, the package has to leverage the installation hooks, namely: \textit{pre-install}, \textit{install}, \textit{post-install}, \textit{preprepare}, \textit{prepare}, \textit{postprepare}, and \textit{prepublish} (deprecated). 
An example is shown in Listing~\ref{lst:js-poc1}. 
%%
%
%
% It is worth mentioning that scientific works discussing malicious packages in npm~\cite{sejfia2022practical,10.1145/3488932.3523262,ohm2023you,ohm2022feasibility} only consider the first three install-hooks mentioned (i.e., \textit{pre-install}, \textit{install}, \textit{post-install}), and not the other ones mentioned in our work.
%
At this stage, attackers have the possibility to directly execute malicious shell commands, or to invoke external scripts which must be included within the package.

% \begin{lstlisting}[label={lst:js-poc1}, caption={Install-time \ac{PoC} implementation for JavaScript using installation hooks in package.json}]
% 	{
% 		"name": "example",
% 		"version": "1.0.0",
% 		...
% 		"scripts": {
% 			@\textbf{"pre-install": "..COMMANDS.."}@,
% 			@\textbf{"install": "..COMMANDS.."}@,
% 			@\textbf{"post-install": "..COMMANDS.."}@ ,
% 			@\textbf{"preprepare": "..COMMANDS.."}@,
% 			@\textbf{"prepare": "..COMMANDS.."}@,
% 			@\textbf{"postprepare": "..COMMANDS.."}@,
% 			@\textbf{"prepublish": "..COMMANDS.."}@
			
% 		},
% 		...
% 	}    
% \end{lstlisting}

\begin{lstlisting}[label={lst:js-poc1}, caption={(I1) Example implementation for JavaScript using installation hooks in package.json}]
	{
		"name": "example",
		"version": "1.0.0",
		... continues ...
		"scripts": {
			@\textbf{"pre-install": "** COMMANDS **"}@
		}
	}    
\end{lstlisting}
% ~\cite{npmjsScriptsDocs}

\noindent\textbf{Recommendation(s).} The \texttt{npm install} command provides the \texttt{-{}-ignore-scripts} option, to avoid the execution of any script during installation~\cite{NpminstallDocs}. If this solution is not viable, it is important to carefully review the content of the installation scripts. This review should be performed for all dependencies being installed, both direct and transitive. However, considering the large number of dependencies that can be required for a single npm package, automation is necessary. Currently, there are no ready-made solutions to address this problem, but academia has started proposing solutions~\cite{10.1145/3488932.3523262, ohm2023you} to mitigate install-time attacks for JavaScript (npm).
Still, works discussing malicious packages in npm~\cite{sejfia2022practical,10.1145/3488932.3523262,ohm2023you,ohm2022feasibility} primarily discuss \textit{pre-install}, \textit{install}, and \textit{post-install} hooks, rather than extensively exploring the abovementioned hooks.

\textit{PHP (composer).}
In PHP, the popular package manager is \textit{Composer}, which supports two types of package distributions: \textit{dist} and \textit{source} packages.
Dist packages are pre-built packages distributed in a binary format. When installing a dist package, Composer skips the build process and directly uses the pre-built package. By default, dist packages are consumed over source packages.
Source packages contain the source code of a package and must be built by the client. The \textit{composer.json} file (equivalent to \textit{package.json} for npm packages) contains the build instructions and offers installation hooks, using the \textit{scripts} property, to execute additional commands within the installation process. 
%This functionality constitutes the basis of the procedure described below.

\noindent\textbf{Procedure.} To achieve \ac{ACE} when installing a package using the \texttt{composer install} command, the following installation hooks can be defined in the \texttt{script} property of the \textit{composer.json} file~\cite{composerScripts}: \textit{pre-install-cmd}, \textit{post-install-cmd}, \textit{pre-autoload-dump}, and \textit{post-autoload-dump}. The implementation is similar to the case of JavaScript (cf. Listing~\ref{lst:js-poc1}). If the package does not include the \textit{composer.lock} file, which records the exact versions of the installed dependencies, the additional installation hooks \textit{pre-update-cmd} and \textit{post-update-cmd} will be executed during the installation (otherwise, they are executed only when \texttt{composer update} is run). 
As for npm packages, attackers may directly insert shell commands in \textit{composer.json}, or invoke external scripts which must be included in the package.

% \begin{lstlisting}[label={lst:php-poc1}, caption={Install-time \ac{PoC} implementation for PHP using installation hooks in composer.json}]
% 	{
% 		"name": "example",
% 		...
% 		"scripts": {
% 			@\textbf{"pre-install-cmd": [
% 				"..COMMANDS.."
% 				]}@,
% 			@\textbf{"post-install-cmd": [
% 				"..COMMANDS.."
% 				]}@,
% 			@\textbf{"pre-update-cmd": [
% 				"..COMMANDS.."
% 				]}@,
% 			@\textbf{"post-update-cmd": [
% 				"..COMMANDS.."
% 				]}@,
% 			@\textbf{"pre-autoload-dump": [
% 				"..COMMANDS.."
% 				]}@,
% 			@\textbf{"post-autoload-dump": [
% 				"..COMMANDS.."
% 				]}@
% 		}
% 	}
% \end{lstlisting}

% \begin{lstlisting}[label={lst:php-poc1}, caption={(I1) \ac{PoC} implementation for PHP using installation hooks in composer.json}]
% 	{
% 		"name": "example",
% 		...
% 		"scripts": {
% 			@\textbf{"pre-install-cmd": [
% 				"..COMMANDS.."
% 				]}@
% 		}
% 	}
% \end{lstlisting}

\noindent\textbf{Recommendation(s).} 
Unlike \texttt{npm install}, the \texttt{composer install} command does not have an option to skip the execution of any script~\cite{getcomposerCommandlineInterface}. As mentioned earlier, the hooks \textit{pre-update-cmd} and \textit{post-update-cmd} are skipped if \textit{composer.lock} exists. The \textit{pre-autoload-dump} and \textit{post-autoload-dump} scripts can also be skipped by using the \textit{-{}-no-autoloader} option. However, there is no built-in way to skip \textit{pre-install-cmd} and \textit{post-install-cmd} when using \texttt{composer install}. Therefore, it is necessary to examine the content of such installation hooks to ensure they do not execute any malicious code. Moreover, it would be important to evaluate the extension to Composer of approaches like~\cite{10.1145/3488932.3523262,ohm2023you}.
%Besides, techniques\todo{countermasures/preventions instead of techniques?} proposed for npm (such as~\cite{10.1145/3488932.3523262,ohm2023you}) could be adapted and applied to Composer as well.
%
Recall that the aforementioned procedure only applies to source packages. Thus, it is crucial to prioritize dist packages for all required dependencies (both direct and transitive) and promote their availability.
% to avoid executing commands at installation time.

\subsubsection*{(I2) Insert code in build script~\cite{ohm2020backstabbers}}
This technique involves the execution of code contained in scripts used by package managers %(e.g., \textit{setup.py} in pip), 
during the installation of 3rd-party dependencies distributed as source code (as they need to be built before usage).

\textit{Python (pip).}
Python packages may be distributed as source or binary distributions. Source distributions (\textit{sdists}) include an installation script, the \textit{setup.py} file, to define the metadata (e.g., name, version), configurations for building and packaging, and additional actions that may be required.
%The \textit{setup.py} file is used in source distributions (\textit{sdists}) for Python projects to define the metadata (e.g., name, version) and configurations for packaging and distribution. 
%Developers may also specify further actions that may be required when installing the package.
The \textit{pip} package manager uses \textit{setup.py} as a source of information to install and manage Python \textit{sdists} packages. 
When running \texttt{pip install}, the \textit{pip} tool executes the \textit{setup.py} file of the package being installed.
%Attackers may exploit the execution of \textit{setup.py} to run arbitrary commands at install time.

\noindent\textbf{Procedures.}
To achieve \ac{ACE} when installing a package through \texttt{pip install}, an attacker can directly add malicious Python commands to the \textit{setup.py} file. An example is shown in Listing~\ref{lst:py-poc1} (malicious instruction in line 4).

\begin{lstlisting}[label={lst:py-poc1}, caption={(I2) Example implementation for Python sdist packages through code in setup.py}]
	from setuptools import setup
	
	# Any Python code will be executed, for example:
	@\textbf{import os; os.system("..COMMANDS..")}@
	setup(name='foo',version='1.0', ...)   
\end{lstlisting}

%\textbf{Procedure 2.}
Another way to achieve \ac{ACE} when installing a package through \texttt{pip install} is to leverage the \texttt{cmdclass} property that allows the customization of the tasks performed.
% during the package installation or build. 
Commonly used cmdclass commands available in setuptools include \texttt{install} and \texttt{build}.
Listing~\ref{lst:py-poc2} demonstrates how to customize the install command class to obtain \ac{ACE} at installation time. First, it is necessary to import the \texttt{install} method from the \texttt{setuptools.command} module (cf. line 2). Then, a subclass of \texttt{install} (\texttt{ExampleClass}, line 4) is created inside the \textit{setup.py} file, that must implement the \texttt{run} method (line 6), which is executed by default at package installation. %An example implementation is shown in Listing~\ref{lst:py-poc2}. 
Malicious code inside the \texttt{run} method is thus executed.

\begin{lstlisting}[label={lst:py-poc2}, caption={(I2) Example implementation for Python sdist packages through cmdclass commands in setup.py}]
	from distutils.core import setup
	@\textbf{from setuptools.command.install import install}@  # Required import
	
	class ExampleClass(install):
		def @\textbf{run(self)}@:
			install.run(self)
			# Any Python code will be executed, for example:
			@\textbf{import os; os.system("**COMMANDS**")}@
	
	setup(name='foo', ..., @\textbf{cmdclass={'install': ExampleClass}}@)
\end{lstlisting}

\noindent\textbf{Recommendation(s).}
Achieving \ac{ACE} primarily depends on consuming source distributions of Python packages for which the installation script is executed during \texttt{pip install}. On the contrary, pre-built packages (i.e., \textit{bdists}) do not include installation scripts, nor execute code during the installation process. The challenge with bdists is that they need to be produced for each target architecture, i.e., a different version of the same Python package has to be built and published for each target architecture possibly willing to install it. Whenever bdists are available, they are the default choice of package managers. Therefore, when selecting a package to install, packages with bdist distributions should be prioritized. The same logic must be applied to all the direct and transitive dependencies as they are transparently installed together with the selected package.
\texttt{pip} allows to ignore sdist dependencies via the option \texttt{-{}-only-binary :all:}~\cite{pypaInstallDocumentation}. Yet, packages without binary distributions will fail to install, possibly preventing the successful outcome of the installation process.
%
%As demonstrated earlier, achieving \ac{ACE} primarily depends on consuming source distributions of Python packages that use the \textit{setup.py} file for installation. However, it is important to note that pre-built packages (i.e., \textit{bdist}) do not include installation scripts (such as \textit{setup.py}), nor execute code during the installation process. The challenge with bdist\todo{change to "build distribution (bdist)"?} releases is that a separate bdist needs to be produced for each target architecture, i.e., a different version of the same Python package has to be build and published for each system possibly willing to install it. When bdist releases are available for packages, they are the default choice of package managers. Therefore, when choosing a package to install, a recommendation is to always prioritize packages that are released through bdists. Such check\todo{which checks? ensuring bdist?} should not only be applied to the package that we want to install, but also to all its dependencies in the entire dependency tree (i.e., all the transitive dependencies).
%For example, \texttt{pip} allows to ignore sdist dependencies via the option \texttt{-{}-only-binary :all:}. Yet, packages without binary distributions will fail to install, possibly preventing the good outcome of the installation process.
%
% \draft{As a side note, pip allows to ignore sdist dependencies via the command \texttt{-{}-only-binary :all:}. Yet, packages without binary distributions will fail to install, possibly preventing the good outcome of the installation process.}
%
If sdists are required (either directly or indirectly), it is crucial to verify whether they include a \textit{setup.py} file and ensure that it does not contain any malicious code.

\textit{Rust (cargo).}
In Rust, the \textit{Cargo.toml} file is used to specify package metadata as well as its direct dependencies. Cargo (i.e., package manager for Rust) uses such a file when running \texttt{cargo install}. To install a package, cargo also builds the package itself as well as all its direct and transitive dependencies. % are fetched, built, and made available for use in the application.
In addition, cargo provides the flexibility to include build scripts within a package that are compiled and executed just before the package is built~\cite{rustlangBuildScripts}. 
By default, the cargo build system will search for a \texttt{build.rs} script in the root directory of the project. This behavior can be overridden by specifying a different path to the build script in the \texttt{build} field of the \texttt{Cargo.toml} configuration file.
Such a feature allows to perform tasks such as building 3rd-party non-Rust code.

% By default, the \texttt{build.rs} script in the root directory of the package is executed. Custom scripts can be configured in the \texttt{build} field of the \texttt{Cargo.toml} file.  %However it involves the execution of arbitrary scripts at installation time.

\noindent\textbf{Procedure.}
To achieve \ac{ACE} when installing a package through \texttt{cargo install}, the attacker  
%The execution of build scripts during the build process introduces a potential attack vector for achieving \ac{ACE} at build-time. 
%If not specified differently in the \texttt{build} field of the \texttt{Cargo.toml} file, the \texttt{cargo build} command will by default search a \texttt{build.rs} script in the  root directory of the project.
%
%An attacker
can include a malicious build script in a package (i.e., a crate) they distribute, that will eventually be executed by the cargo build system. Listing~\ref{lst:rust-poc} shows how to trigger the execution of the commands in line 5 using the \textit{build.rs} script.
%When a victim includes this malicious crate as a dependency in their own application and builds it using Cargo, the malicious build script will be executed as part of the build process. This enables the attacker to inject and execute their malicious code on the victim's system during the build.

\begin{lstlisting}[label={lst:rust-poc}, caption={(I2) Example implementation for Rust leveraging \textit{build.rs}}]
	use std::process::Command;
	
	fn main() {
		# Any arbitrary Rust code can be executed, for example:
		@\textbf{let output =  Command::new("sh")
			.arg("-c")
			.arg("**COMMANDS**")
			.output();}@
	}
\end{lstlisting}

% \begin{figure}[!ht]
	%     \captionsetup{type=lstlisting}
	%     \begin{sublstlisting}{\linewidth}
		%     \begin{lstlisting}
			% [package]
			% name = "example"
			% version = "0.1.0"
			% ...
			
			% @\textbf{[build-dependencies]}@
			%     \end{lstlisting}
		%     \caption{Content of the \textit{Cargo.toml} file for the project}
		%     \label{lst:file1}
		%     \end{sublstlisting}
	
	% \begin{sublstlisting}{\linewidth}
		% \begin{lstlisting}
			% use std::process::Command;
			
			% fn main() {
				%     # Any arbitrary Rust code will be executed, for example:
				%     @\textbf{let output =  Command::new("sh")
					%     .arg("-c")
					%     .arg("..COMMAND..")
					%     .output();}@
				% }
			%     \end{lstlisting}
		%     \caption{Content of \textit{extconf.rb} file}
		%     \label{lst:file2}
		%     \end{sublstlisting}
	%     \caption{Build-time \ac{PoC} implementation for Rust cargo through code leveraging build dependencies}
	%     \label{lst:rust-poc}
	%     \end{figure}

\noindent\textbf{Recommendation(s).} To ensure the security of consumed 3rd-party dependencies, it is recommended to check for the presence of build scripts, i.e., \textit{build.rs} or the one specified within the \texttt{build} field in the \textit{Cargo.toml} file. This applies to both direct and transitive dependencies. 
Reviewing the content of these scripts is crucial to identify and mitigate potential malicious code.

% \subsubsection*{(Out of Scope) Vulnerabilities in the toolchain}
% \draft{Go case of CVE-2023}

\subsubsection*{(I3) Run code in build extension(s)}
This technique involves executing extensions of 3rd-party dependencies that are necessary for their build process. 

\textit{Ruby (gem).}
The \textit{.gemspec} file contains the metadata and dependencies for a Ruby package (i.e., a gem). Such a file is used by the RubyGems package manager to install, build, and distribute a package.
Gems may include extensions that are built and executed at installation time (i.e., when running \texttt{gem install})~\cite{rubygemsGemsWith}. Note that extensions are also executed when manually building a 3rd-party dependency through the \texttt{gem build} command. 

\noindent\textbf{Procedure.}
To achieve \ac{ACE} when installing a package through \texttt{gem install}, two files needs to be manipulated: the \textit{.gemspec} and extension files.
%Such functionality can be exploited by malicious dependencies to achieve \ac{ACE} as shown in Listing~\ref{lst:ruby-poc}.
%While for JavaScript, PHP, and Python, the minimum number of files involved in the process of achieving \ac{ACE} at install-time is just one, in the case of Ruby two files are involved (i.e., the \textit{gemspec} file and the one related to the extension).
As shown in Listing~\ref{lst:ruby-poc}, an attacker may define a build extension in \textit{.gemspec} (line 5 in Listing~\ref{lst:file1}) and include it in the gem. Malicious code in the extension file will be executed (line 4 in Listing~\ref{lst:file2}).

\begin{figure}[!ht]
	\captionsetup{type=lstlisting, position=above}
	\caption{(I3) Example implementation for Ruby gems leveraging build extensions}
	\begin{sublstlisting}{\linewidth}
		\begin{lstlisting}
			Gem::Specification.new do |s|
			s.name        = "example"
			s.version     = "1.0.0"
			... continues ...
			@\textbf{s.extensions  = ["extconf.rb"]}@
			end\end{lstlisting}
		\caption{Content of the \textit{.gemspec} file for the project}
		\label{lst:file1}
	\end{sublstlisting}
	\begin{sublstlisting}{\linewidth}
		\begin{lstlisting}
			require "mkmf"
			
			# Any arbitrary Ruby code will be executed, e.g.:
			@\textbf{exec("**COMMANDS**")}@
			
			# Needed to finish the extension without errors
			create_makefile("")\end{lstlisting}
		\caption{Content of \textit{extconf.rb} file}
		\label{lst:file2}
	\end{sublstlisting}
	
	\label{lst:ruby-poc}
\end{figure}

\noindent\textbf{Recommendation(s).}
The \texttt{gem install} command do not provide an option to ignore extensions. 
In order to prevent such a scenario, gems should be checked to verify if they declare extensions in  \textit{.gemspec}. If present, extensions should be analyzed to verify that they do not contain malicious code. Such a review has to be performed both for direct and transitive dependencies.

\subsection{(R) Runtime Execution}\label{sec:runtime-exec}

Malicious code can be executed at runtime through various techniques. %Attackers aim to have their code executed by victims, and 
We identify four scenarios (cf. Table~\ref{tab:ace-techniques}) where this is more likely to occur: compromising the code executed during the import of external modules, manipulating popular methods, manipulating constructor methods, or leveraging build plugins. %tampering (or monkey patching) commonly used methods, 
%Table~\ref{tab:lang-comparison} maps them to the selected languages.  

\subsubsection*{(R1) Insert code in methods/scripts executed when importing a module~\cite{ohm2020backstabbers}}

%In this section, we delve into a sub-technique\todo{why is it sub-technique?} associated with 
% Runtime execution of 3rd-party code may occur when developers import an external module. In fact, certain programming languages execute code when an import statement is processed, even before the code from the imported module is actually used (for Javascript, Python, and Ruby even if the imported module is never used).
% %enabling specific actions to take place at this stage. 
% Below, we examine the procedures for programming languages subject to this technique.%\todo{this paragraph can be shortened} 
Certain programming languages execute code when an import statement is processed, even before the code from the imported module is actually used (for Javascript, Python, and Ruby even if the imported module is never used). Thus, runtime execution of 3rd-party code may occur when developers import an external module.
%enabling specific actions to take place at this stage. 
Below, we examine the procedures for programming languages susceptible to this technique.%\todo{this paragraph can be shortened} 

\textit{JavaScript (npm).}
In Node.js, the \texttt{main} attribute~\cite{nodejsModulesPackages} in the \textit{package.json} file determines the entry point script (e.g., \texttt{index.js}) for a package when it is imported into an application (e.g., using \texttt{require}). Attackers can inject malicious code into the entry point script such that it will be executed at runtime when the 3rd-party module is imported, thus achieving \ac{ACE}. % at import-time.

\textit{Python (pip).}
Regular packages in Python are typically implemented as a directory containing an \texttt{\_\_init\_\_.py} file~\cite{pythonImportSystem}. Attackers may insert malicious code inside these files since their execution is triggered implicitly when packages are imported using the \texttt{import} statement.

\textit{Ruby (gem).}
%
%In addition to installation time, Ruby automatically executes code at runtime upon a \texttt{require `dependency'} statement. There are two ways this can be triggered: First, when a Ruby dependency is imported via the \texttt{require `dependency'} statement, the Ruby code of the dependency (\texttt{lib/ dependency.rb} file) is automatically executed. 
% \draft{Ruby automatically executes code at runtime upon a \texttt{require} statement in two cases.
%
% First, when a Ruby dependency is imported via the \texttt{require} statement, the Ruby code of the dependency (\texttt{lib/dependency.rb} file) is automatically executed. 
%
% when a C extension module is loaded and it defines a the \texttt{Init\_<extension name>()} method.
% The \texttt{Init\_<extension name>()} method is defined in the C extension's source code and serves as the entry point for initializing the extension module. This method is invoked when the module is loaded into the Ruby interpreter.}
Ruby automatically executes code at runtime upon a \texttt{require}, \texttt{require\_relative} or \texttt{load} statement. Thus, attackers may insert malicious code within the \texttt{.rb} file from which the module is imported.

% NB: next paragraph commented and moved the c-extensions to future work
%Furthermore, attackers may leverage C extension modules in Ruby to further hide the malicious code. To do so, the C extension has to define the \texttt{Init\_<extension-name>()} method, which specifies the tasks to be performed when initializing the extension module upon requiring the dependency. Note that in the case of C extension modules, attackers need to compile the code for the specific architecture of the victim machine. 

% In Ruby it is also possible to provide C or Java extensions in a package~\cite{rubygemsGemsWith}. At runtime, if a dependency uses an extension via the \texttt{require} statement, the \texttt{Init\_<extension name>()} method of the extension is automatically executed. Thus, it is possible to execute code located in extensions, by requiring a dependency. 
%As in JavaScript and Python, note that Ruby allows to require dependencies even if they are never used after being imported.\todo{say this for other languages too (except go?)} 
%Second, it is possible to provide C or Java extensions in a Ruby package~\cite{rubygemsGemsWith}. 
%During runtime, if a dependency uses an extension via the \texttt{require 'extension'} statement, the \texttt{Init\_<extension name>()} method
%of the extension is also automatically executed. Thus, it is possible to execute code located in extensions, by requiring a dependency. Note that, Ruby allows to require dependencies even if they are never used %after being imported.\todo{say this for other languages too (except go?)} 

\textit{Go (go).}
There are several precautions taken by Go against software supply chain attacks~\cite{goSupplyChain}.
% cite https://go.dev/blog/supply-chain
It does not run any code on installation and there are no installation hooks, preventing the techniques of Section~\ref{sec:install-time}. 
%Second, it does not run any code at build times, even if this was recently contradicted by a CVE~\cite{CVE-2023-29405}.
%Second, it does not run any code at build time.\todo{theres a CVE here (commented) but I dont know if its worthwhile to mention it}
%
Moreover, unlike other programming languages analyzed in this work, Go lacks a centralized package repository. Instead, packages are directly downloaded from their source code repositories. 
In Go, dependencies can execute code upon import in two ways: \emph{(i)} by defining an \texttt{init()} method~\cite{goEffectiveProgramming,goBlogPost}, and \emph{(ii)} by initializing a variable with an anonymous function~\cite{goBlogPost}.
This allows the code of a 3rd-party dependency to be executed with higher order of precedence, before the code of the importing application. Additionally, Go automatically removes unused dependencies by default, but prefixing the import with an underscore symbol (e.g., \texttt{\_ "foo"}), retains even unused dependencies.
Attackers can exploit these techniques to craft a malicious dependency that achieves \ac{ACE} at runtime during its import~\cite{michenriksenFindingEvil}.

\subsubsection*{(R2) Insert code in commonly-used methods}
Attackers may target popular methods within 3rd-party dependencies that they distribute to downstream users. These methods are attractive because they are commonly used, increasing the likelihood that downstream users will invoke and rely on them. 
%
% For example, if a malicious package is typosquatting a popular one (e.g., \texttt{pandas} for Python), malicious code would be included in widely-used functions (e.g., \texttt{read\_csv()}).\textbf{SP: can we assume typosquatting is self-explanatory?}
For example, this technique is used in the Java malicious package \texttt{com.github.codingandcoding: servlet-api-3.2.0}~\cite{dasfreakBackstabbersKnife,sonatypeSonatypeStops}, where the malicious code is contained in the \texttt{doGet()} method of the \texttt{HttpServlet} class.
%\textbf{I think this additional info is nice, at the same time it makes me think it's missing for the other techniques. I have mixed feelings.}

\subsubsection*{(R3) Insert code in constructor methods (of popular classes)}
%As a specific instance of the previous technique (R2)\todo{if it is an instance of R2, should we list it separate?}, 
Attackers may target constructor methods as suitable places to insert malicious code. In fact, constructors are called during object instantiation, making them potential entry points for malicious activities. For instance, in a typosquatted version of the Python \texttt{pandas} package~\cite{ohm2020backstabbers,ladisa2022taxonomy}, malicious code could be included in the constructor methods of the widely-used class \texttt{Dataframe()}.

Regarding Java, it is also noteworthy to mention that attackers may not only include malicious code in constructor methods but also in instance and static initializers~\cite{oracleChapterClasses}. Instance initializers are executed every time a class instance is created~\cite{JavaSpec12_4_2}. Static initializers are automatically executed (just once) when a class is initialized, i.e., when a class instance is created, a static method of the class is called, or a static field of the class is assigned or used (excluding constants)~\cite{JavaSpec12_4_2}.

% \subsubsection*{(R4) Replace popular methods using monkey-patching~\cite{ohm2020backstabbers} or function hooking}
% Malicious 3rd-party packages may execute malicious code at runtime by replacing methods that are likely used by downstream users~\cite{ohm2020backstabbers}.
% In dynamically-typed languages (e.g., Python, JavaScript, Ruby), this is possible by using monkey patching~\cite{10.1145/3098954.3120928}.
% In statically-typed languages (e.g., Rust, Go), this is possible using function/API hooking~\cite{bouMonkeyPatching}.
% %
% In the Java programming language, attackers can inject malicious code at runtime by modifying the bytecode of a trusted Java class file that is used by downstream users or by abusing the classloading mechanism~\cite{williams2009enterprise}.

\subsubsection*{(R4) Run code of 3rd-party dependency as build plugin}
This technique consists of running a 3rd-party dependency as a plugin %of the build tool employed when building downstream project.
within the build of a downstream project.

\textit{Java (mvn).}
%In Maven, plugins enable developers to customize and extend the functionality of the build process. Plugins are defined in the project's Maven configuration file (\textit{pom.xml}) and they allow to specify tasks to automate various build processes (e.g., compiling code, running tests, packaging artifacts).
In Maven, plugins enable developers to define tasks to be performed. Plugins are defined in the Maven configuration file of the project (i.e., \textit{pom.xml}) and they can be bound to phases of the build process (e.g., compile, test, package), thus customizing and extending its functionality.
Attackers can inject malicious code into Maven plugins as a means to achieve \ac{ACE} during the build process or to infect the built artifact to spread malicious code. For example, this technique is exploited by the malicious Java package \texttt{com.github.codingandcoding: maven-compiler-plugin-3.9.0}~\cite{sonatypeSonatypeStops}.

\smallskip\noindent
\textbf{Remarks.}
%
%When using 3rd-party dependencies in tests, they can employ the techniques described for the runtime cases to achieve \ac{ACE} (e.g., import-time case).\todo{not clear sentence. also we need to explain why this is out of scope too.}
%
When using 3rd-party dependencies in the tests of a project, the same techniques described in Section~\ref{sec:runtime-exec} to achieve \ac{ACE} apply. 
Additionally, attackers may ship packages with malicious code in the test routines~\cite{ohm2020backstabbers}. This is out of scope in our work, since we focus on the execution of malicious code contained in 3rd-party dependencies during the installation, build, test, or runtime of downstream projects. 
As a result, malicious code in tests of 3rd-party dependencies is not executed as those tests are not executed in any stage of the lifecycle of the downstream project.
%As a result, the tests of 3rd-party dependencies are not executed.  

From the attacker's viewpoint, install-time techniques provide greater advantages than runtime techniques. In the former, victims simply need to install the package to initiate the malicious code execution, which might explain the prevalence of malicious packages executing code at install-time~\cite{ohm2020backstabbers,zahan2022weak}. Conversely, with runtime techniques, victims not only have to install the third-party dependency but also actively trigger the carrier of the malicious code (e.g., invoking the constructor as demonstrated in the R2 technique).

%% file: sections/rq2.tex
\section{RQ2: Evasion Techniques}\label{sec:rq2}

In this section, we present various techniques that aim to make the detection of malicious source code more challenging. We classify these techniques into the three categories of software obfuscation — namely, \textit{data obfuscation}, \textit{static code transformation}, and \textit{dynamic code transformation} — as proposed by Schrittwieser et al.~\cite{10.1145/2886012}.

The covered techniques include both those that have been observed in real-world attacks (e.g., in \ac{BKC}~\cite{dasfreakBackstabbersKnife}) and those that are theoretically viable based on state-of-the-art techniques in code obfuscation~\cite{10.1145/2886012,1027797, xu2017secure,collberg1997taxonomy}. Such techniques can also be combined and used in conjunction to further challenge the detection and analysis of malicious code.

While we strive for comprehensiveness, it is important to note that this list is not exhaustive and may evolve over time as new techniques emerge.

%It is important to note that the techniques presented in this section should not be viewed as isolated operations carried out by attackers. Instead, they can be combined and used in conjunction to further challenge the detection and analysis of malicious code.

%However, we believe that understanding these techniques can enhance defense strategies and develop countermeasures to mitigate the risks associated with malicious code in software supply chains.

%\todo{I shortened this part, prev. version is in comments}

\subsection{Data Obfuscation}

Malicious code often incorporates hard-coded strings that serve various purposes for attackers. These strings can include, e.g., URLs or IP addresses that point to attacker-controlled servers, shell commands, or paths to sensitive files~\cite{10.5555/2181153,10.1145/3560835.3564548}.
Since the analysis of these strings can provide insights into the attacker's infrastructure, techniques, and intended actions, attackers often employ obfuscation techniques to evade detection and conceal information that could lead to their identity.
Thus, in this category, we encompass techniques that alter the way static data (i.e., strings) is stored within source code to conceal it from analysis and enhance its complexity~\cite{collberg1997taxonomy,10.1145/2886012}.

\textit{Encoding.} 
One commonly used technique involves encoding strings in non-human-readable formats (e.g., base64, hex) and decoding them at runtime~\cite{ohm2020backstabbers}. From an attacker's perspective, this technique is easy to implement and can effectively evade detection through simple string scanning techniques that look for sensitive words (e.g., \texttt{bash}). However, strings encoded in common formats are relatively easy to de-obfuscate.

\textit{Compression.} Attackers can employ compression algorithms (e.g., \textit{gzip}) to obfuscate strings~\cite{ohm2020backstabbers}. The compressed strings would then be decompressed at runtime to retrieve their original content.  

\textit{Encryption.}
Attackers may encrypt strings using common algorithms (e.g., AES) and decrypt them at runtime~\cite{ohm2020backstabbers}. This approach increases the complexity of the analysis process, as the encrypted strings are not directly readable and require the decryption key to reveal their original content. The key used for decryption can be handled in different ways. In some cases, it may be stored locally within the package itself, making it relatively easy for security analysts to extract and decrypt the strings. To mitigate this risk, attackers may choose to retrieve the decryption key from a remote server at runtime. This approach adds an additional layer of complexity, as the decryption key is not readily available in the package. By relying on a remote server, attackers can dynamically control the availability and distribution of the decryption key, making it more challenging for security analysts to access and decrypt the strings.

\textit{Binary Arrays.} 
A more advanced technique is to represent strings in a binary form and store them in binary arrays. Binary arrays provide a convenient way to store and manipulate binary data in a programmatic way. By leveraging binary arrays, attackers can perform operations such as bitwise operations, XOR operations, or custom encoding schemes to further obfuscate the strings~\cite{7583913}. To effectively detect and analyze such procedures, security practitioners need to employ advanced techniques capable of handling binary data and apply reverse engineering methods.

%\textit{ Compression.} 
%To decrease the readability of strings, attackers may employ compression techniques (e.g., using algorithms like gzip) to compress the strings. The compressed strings would then be decompressed at runtime to retrieve their original content.\todo{compression is also a form of encoding}

\textit{Reordering of Data.} 
Another simple and effective technique is to split data into multiple pieces and re-aggregate them at runtime~\cite{collberg1997taxonomy,10.1145/2886012}. Attackers can manipulate strings, such as URLs or shell commands, by splitting them into multiple chunks (e.g., assigned to multiple variables) and then concatenating them before providing them to the respective APIs or functions~\cite{ohm2020backstabbers}. By breaking down the strings into smaller fragments and reassembling them dynamically, attackers can evade straightforward pattern matching techniques~\cite{10.1145/2886012} that rely on static scanning of strings for detection. %For example, this technique
Security analysts need to employ more advanced detection that can handle dynamic string manipulation to effectively mitigate such attacks.

%\textit{ Custom Transformations.}
%As described below, the techniques presented can be combined to create custom transformation techniques. Attackers may alter the order of characters, encode specific parts of the input strings, and employ various other transformations. This introduces an additional layer of complexity, requiring analysts to reverse engineer the transformation technique used in order to recover the original content of the obfuscated string.\todo{this part can be removed, we already mention that tchniques can be combined.}

\subsection{\textbf{Source Code Transformations}}

To make it challenging for security analysts to understand the purpose and inner workings of malware, attackers employ various techniques to obfuscate the source code.
The obfuscation process aims to create a complex and convoluted code structure that hinders reverse engineering and analysis.

\subsubsection*{\textbf{Static Code Transformations}}

In this category, we cover techniques involving the transformation of source code that do not necessitate additional runtime modifications for execution~\cite{10.1145/2886012}.

\textit{Renaming Identifiers.} 
To decrease the readability of code, attackers may employ the technique of renaming identifiers such as variable and function names~\cite{7583913}. By changing the names of these identifiers to arbitrary or nonsensical values, attackers make it challenging for analysts to understand the purpose and functionality of the code.
An example of such a technique is shown in Listing~\ref{lst:maratlib}.

\begin{lstlisting}[label={lst:maratlib}, caption={Obfuscated code in the \textit{setup.py} of the package \textit{maratlib-0.2}~\cite{dasfreakBackstabbersKnife}}]
	import sys
	l1l_cringe_ = sys.version_info [0] == 2
	l1l1l_cringe_ = 2048
	l11_cringe_ = 7
	def l111_cringe_ (l1ll_cringe_):
		global l11l1_cringe_
		l11l_cringe_ = ord (l1ll_cringe_ [-1])
		ll_cringe_ = l1ll_cringe_ [:-1]
		l1l1_cringe_ = l11l_cringe_ % len (ll_cringe_)
		l1_cringe_ = ll_cringe_ [:l1l1_cringe_] + ll_cringe_ [l1l1_cringe_:]
	... continues ...
\end{lstlisting}

\textit{Dead/Useless Code Insertion.} 
Attackers may insert dead or useless code into their malicious code (e.g., \texttt{mplatlib-1.0}~\cite{dasfreakBackstabbersKnife}) to deceive security analysts during the reverse engineering process and to make the latter more time-consuming~\cite{5633410,7583913}.

Dead code refers to portions of code that are intentionally added but never executed during the runtime of the program. It serves no functional purpose and may contain instructions or logic that are irrelevant to the actual operation of the malware.

Useless code, on the other hand, refers to code that may be syntactically correct and executable but serves no meaningful purpose in terms of the malware's functionality. As an example, it may consist of redundant or duplicated code, excessive comments, or superfluous operations.

\textit{Split Code into Multiple Files.} 
Attackers may employ the technique of splitting malicious code into multiple files within the same package to obfuscate their activities and make detection and analysis more challenging. By distributing the malicious code across multiple files, it becomes harder for security analysts to identify and understand the complete scope of the malicious functionality.

Similarly, attackers employ the technique of leveraging second-stage payload(s)~\cite{ohm2020backstabbers}. 
Instead of directly including the malicious code within the package, attackers only provide the initial code for fetching the actual malicious payload from external sources.
By utilizing a remote server, attackers gain dynamic control over the availability and distribution of the second-stage payload, which increases the difficulty for security analysts to access the actual malicious code. This approach makes more complex the analysis process, as the initial code alone does not reveal the full extent of the malicious activities. 

\textit{Hide Code into Dependency Tree.} 
Another obfuscation technique used by attackers is the hiding of malicious code within the dependency tree of software packages. This technique involves including the actual malicious code in either the direct dependencies (e.g., as happened in the case of \texttt{event-stream}~\cite{mediumCompromisedPackage}) or the transitive dependencies (which is more effective) of the final package that will be distributed to downstream users in an \ac{OSS} supply chain attack. Where to place the malicious code within the actual malicious dependency is explored in Section~\ref{sec:rq1}.

By embedding the malicious code deep within the dependency tree, attackers attempt to evade detection, as security scanning tools and analysts may have difficulties in thoroughly scanning every single dependency.

\textit{Split Code into Multiple Dependencies.} 
Instead of placing all the malicious code in a single package, attackers may distribute it across multiple dependencies within the package ecosystem. This approach aims to make it more difficult to detect and analyze the malicious code since it is scattered across different packages. To the best of our knowledge, this technique has not been observed (yet) in real-world attacks.

In theory, a comprehensive scan of packages and their relationships within package repositories could potentially uncover such technique. In practice, this is challenging due to the vast number of packages and the continuous influx of new versions and updates.

\textit{ Visual Deception.}
Attackers may employ visual deception techniques to make manual code review more challenging. One such technique involves adding excessive spaces, tabs, or other forms of whitespace to the code~\cite{phylumVisualDeception}.
Compared with the other obfuscation techniques, this one capitalizes on human visual processing, as the excessive spaces can make it difficult for reviewers to spot anomalies or suspicious code patterns. 

Another technique involves the use of Unicode homoglyphs or control characters~\cite{boucher2021trojan}.
The intent is to visually camouflage the malicious code, making it blend in with the surrounding code and potentially bypass casual inspection\footnote{This issue has started being addressed by some IDEs and compilers}.

\textit{Polyglot Malwares and In-Line Assembly.}
Polyglot malwares employ multiple programming languages within a single malware instance. By combining different programming languages, attackers can exploit the unique characteristics of each language, thereby increasing the complexity of their malware.
Moreover, certain programming languages allow the inclusion of assembly instructions directly in the source code. In-line assembly provides a means for direct interaction with system components and precise control over system resources.
To the best of our knowledge, also this technique has not been observed (yet) in real-world attacks.

Effectively analyzing and detecting the behavior of malware that employs multiple languages or incorporates in-line assembly requires advanced skills and expertise from security analysts and researchers. It demands a deep understanding of the intricacies of various programming languages and the low-level operations of the underlying system. Furthermore, traditional security mechanisms and tools may struggle to cope with the complexity introduced by polyglot malwares and in-line assembly.

\subsubsection*{\textbf{Dynamic Code Transformations}}

In this category, we encompass techniques that involve transforming the code at runtime before its execution~\cite{10.1145/2886012}. These techniques are not detectable through static analyzers.

\textit{Encoding, Compression and Encryption.} 
Similar to strings, attackers can transform the malicious code itself using encoding, compression, or encryption techniques~\cite{ohm2020backstabbers}. 
%to make it appear in a non-human-readable format. 
At runtime, the code is then decoded, decompressed or decrypted back to its original form before being executed. This is similar to the concept of \textit{packing} for binary malwares~\cite{10.1145/2886012,7163053}.

%To reverse the encoded, compressed or encrypted source code, security analysts need to identify the specific algorithm used and apply the corresponding decoding, decompression or decryption process. This task requires specialized knowledge and tools to identify the obfuscation technique and apply the appropriate reversal method. 
%In the case of encrypting the source code, similar considerations made for the case of strings apply regarding the use of decryption keys. Thus, when the source code is encrypted, a decryption key is required to reverse the encryption and obtain the original code.

\textit{Steganography.}
Steganography consists of hiding information by embedding it within other data, such as images, audio files, or even seemingly harmless text files. Attackers may employ steganography as a technique to conceal malicious code within innocuous-looking files, as it happened with the case of the package \texttt{apicolor-1.2.4} in PyPI~\cite{checkpointCheckPoint}.

\textit{Dynamic Code Modification.}
Malicious 3rd-party dependencies can manipulate the behavior of commonly used methods by developers (e.g., built-in functions of a language like \texttt{System.out. println} in Java~\cite{geekexplainsTrickyStatic})  before their execution~\cite{10.1145/2886012}. This tactic not only hides the invocation of malicious behavior within apparently harmless calls in the downstream application but also heightens the likelihood of such malicious code being triggered by the victim.

In dynamically typed languages (e.g., Python, JavaScript, Ruby), this functionality is already supported through monkey patching~\cite{10.1145/3098954.3120928}. Listing~\ref{lst:pyth-monkeypatch} illustrates an instance of monkey-patching in Python, where the built-in function \texttt{print} is tampered with to execute shell command(s).
In statically typed languages (e.g., Rust, Go), achieving this goal can be less straightforward and one possible way is through function/API hooking~\cite{bouMonkeyPatching}.
In the Java programming language, attackers can insert malicious code during runtime by altering the bytecode of a trusted Java class file that is utilized by downstream users. This could involve exploiting the Java instrumentation API or manipulating the classloading mechanism~\cite{williams2009enterprise}.

\begin{lstlisting}[label={lst:pyth-monkeypatch}, caption={Example of monkey-patching the built-in function \texttt{print} in Python to make it execute shell command(s)}]
	import os, builtins

	original_print = print
	def hacked_print(self):
		original_print(self)
		@\textbf{os.system("..COMMANDS..")}@
	builtins.print = hacked_print
\end{lstlisting}

% \subsubsection*{(R4) Replace popular methods using monkey-patching~\cite{ohm2020backstabbers} or function hooking}
% Malicious 3rd-party packages may execute malicious code at runtime by replacing methods that are likely used by downstream users~\cite{ohm2020backstabbers}.
% In dynamically-typed languages (e.g., Python, JavaScript, Ruby), this is possible by using monkey patching~\cite{10.1145/3098954.3120928}.
% In statically-typed languages (e.g., Rust, Go), this is possible using function/API hooking~\cite{bouMonkeyPatching}.
% %
% In the Java programming language, attackers can inject malicious code at runtime by modifying the bytecode of a trusted Java class file that is used by downstream users or by abusing the classloading mechanism~\cite{williams2009enterprise}.

\smallskip\noindent
\textbf{Remarks on warning suppression.}
% \subsubsection*{ \textbf{Suppress Warnings}}
To avoid alerting the victim or halting program execution due to exceptions, malware developers often employ a technique where they enclose the malicious code within a try block and associate it with an empty catch block. This technique is used to suppress any exception that may be thrown during the execution of the malicious code~\cite{10.1145/3560835.3564548}.
In programming languages like Python and Java, the try-catch construct allows developers to handle exceptions gracefully and perform appropriate actions when an exception occurs. However, in the case of malware, the catch block is deliberately left empty, effectively silencing any exceptions that may occur.

By using an empty catch block, the malware ensures that any exceptions raised during its execution are not propagated or displayed to the user. This helps maintain stealth and prevents any error messages or abnormal program behavior that may alert the victim or raise suspicion.

From a security standpoint, the presence of empty catch blocks in code should raise suspicion and indicate potentially malicious activities. It is important for security analysts and developers to be vigilant and thoroughly analyze code for such suspicious constructs during code reviews and security assessments.

%% file: sections/relwork.tex
\section{Related Work}\label{sec:related-works}

In this section, we present related works that focus on the security aspect of package managers in the context of \ac{OSS} supply chain attacks.

Ohm et al.~\cite{ohm2020backstabbers} analyze OSS supply chain attacks, investigating malicious packages in npm, PyPI, and RubyGems. They develop an attack tree, outlining techniques used in known malicious packages to trigger execution in software life cycle phases. Our work goes beyond existing malicious examples by investigating additional potential techniques and analyzing a wider range of programming languages.

%explores both targeted and similar ecosystems to comprehensively understand execution triggers across different ecosystems, anticipating attackers and improving security.

Ladisa et al.~\cite{ladisa2022taxonomy} build upon Ohm et al.'s work~\cite{ohm2020backstabbers} and conduct a systematic study to create a comprehensive taxonomy of how attackers inject malicious code into \ac{OSS}. However, this study does not address the actual malicious content of packages.  
Our work complements this by analyzing the different execution and evasion techniques leveraged by malicious packages. 
%Our work complements this by analyzing the techniques of how attacker can inject malicious code and enable its execution. 
%e package managers' features in popular ecosystems to induce the execution of injected malicious code.

Okafor et al.~\cite{10.1145/3560835.3564556} provide a systematic analysis of secure software supply chain patterns. In particular, they present a framework consisting of four stages of a software supply chain attack (i.e., compromise, alteration, propagation, and exploitation) and introduce three fundamental security properties (i.e., transparency, validity, and separation). Following their terminology, our work aims to understand the techniques used by attackers on package managers, in the compromise stage of an \ac{OSS} supply chain attack.

Zimmerman et al. \cite{236368} examine the threats and risks faced by users of the npm ecosystem by investigating package dependencies, maintainers, and publicly reported security issues. In a similar fashion, Bagmar et al. \cite{DBLP:journals/corr/abs-2102-06301} conduct a related study on the PyPI ecosystem. 
In contrast to these works, our work focuses on analyzing how dependencies can achieve code execution, not limited to npm or PyPI but also encompassing ecosystems with similar features. By exploring such mechanisms, we aim to provide a broader understanding of security risks across different ecosystems and provide runnable examples.

Duan et al.~\cite{duan2020towards} examine the functionalities of npm, PyPI, and RubyGems, proposing a framework to assess functional and security features and detect malicious packages using program analysis. While their framework has a broad scope (e.g., covering authentication, signing features), our work has a practical approach on package manager features that can lead to \ac{ACE}. Additionally, we extend our analysis to include other package managers (e.g., Composer, Cargo).

Wyss et al.~\cite{10.1145/3488932.3523262} investigate the potential exploitation of install-time features in npm by attackers, and propose Latch as a solution. While they specifically focus on the \textit{pre-install}, \textit{install}, and \textit{post-install} hooks, our work extends the analysis to identify additional installation hooks provided by npm. Furthermore, we explore similar attack vectors in other ecosystems beyond npm, broadening the scope of our investigation.

Zahan et al.~\cite{zahan2022weak} focus on npm, identifying six security weakness signals in software supply chain attacks. They analyze packages at scale to measure these signals and survey 470 package developers to assess their significance. While we share a focus on package manager security, our offensive approach explores how attackers exploit package managers across ecosystems for code execution, studying also evasion techniques.

% {\color{blue} TODO.
	
	% Cappos et al. \cite{10.1145/1455770.1455841,packagemanagementsecurity} identify possible
	% attack vectors related to a lack of proper signature management at the level of
	% packages and their metadata, some of which we considered below \emph{Distribute
		% malicious version of legitimate package}.

	% Wyss et al.~\cite{10.1145/3488932.3523262}. Similarly, Ohm et al.~\cite{ohm2023you}.
	
	% Zahan et al.~\cite{zahan2022weak}
	
	% Gu et al.~\cite{investigatingPackage}

%% file: sections/conclusions.tex
\section{Conclusions}\label{sec:conclusion}

% \draft{
% In this work, we have presented 
% how it is possible to achieve ACE via malicious packages when they are integrated as 3rd-party dependencies.
% %
% In particular, we have posed two research questions, and showed 
% several strategies to achieve ACE in downstream packages
% at install-time (also leveraging the behavior of famous package managers) and runtime, together with possible evasion techniques that malicious actors can employ not to be detected.
% %
% We performed our analysis for 7 ecosystems, and identified 3 install-time and 5 runtime strategies for achieving ACE, and studied 3 classes of evasion techniques.
% %
% Finally, we published our proof-of-concepts implementations to help researchers and practitioners further develop their safeguards.

% As future works, we plan to further extend our analysis to other ecosystems, and to further develop possible recommendations and safeguards.
% }

In this work, we analyze 7 ecosystems and identify 7 different techniques usable by malicious 3rd-party dependencies to achieve \ac{ACE} in downstream projects
at install-time and runtime. Example implementations are provided to aid researchers and analysts in developing countermeasures. We also investigate evasion techniques employed by attackers to challenge detection.
%
% We performed our analysis for 7 ecosystems, and identified 3 install-time and 5 runtime strategies for achieving ACE, and studied 3 classes of evasion techniques.
%

In future work, we plan to extend our analysis to other ecosystems, develop recommendations for downstream users and security analysts, and conduct a study on developers' experiences when installing and using open-source packages.
Moreover, we would like to further develop our analysis of programming languages in hybrid scenarios, like the usage of C extensions in Python and Ruby or the usage of the Java Native Interfaces in Java.
Finally, it is relevant to perform a large-scale analysis that aims to estimate how significant the different techniques of \ac{ACE} are for each ecosystem or the practical implications of using some of the countermeasures (e.g., how many installations fail in Python using the \texttt{--only-binary :all:} option).

% As future works, we plan to further extend our analysis to other ecosystems, and to further develop possible recommendations and safeguards.